\documentclass{jaa}

\usepackage{graphicx}
\usepackage{xspace}
\usepackage{epsf,url,wrapfig ,fancyhdr,multirow,lscape,appendix,rotating, setspace,amsmath,amssymb}
\usepackage{graphics,epsfig,url}
\usepackage{graphicx}
\usepackage{gensymb}
\usepackage{amssymb}
\usepackage{xcolor}
\usepackage{color}
\usepackage{lipsum}
\usepackage{textcomp}

\newcommand{\astrosat}{\textit{AstroSat}~}

\newcommand{\swift}{\textit{Swift}}
\newcommand{\suzaku}{\textit{Suzaku}~}

\newcommand{\fermi}{\textit{Fermi}\xspace}

\newcommand{\source}{OAO~1657--415~}

\newcommand{\fluxcgs}{erg~cm$^{-2}$~s$^{-1}$}

\begin{document}\sloppy

\title{{\em AstroSat} observations of eclipsing high mass X-ray binary pulsar OAO~1657--415}

\author{Gaurava K. Jaisawal\textsuperscript{1,*}, Sachindra Naik\textsuperscript{2}, Prahlad R. Epili\textsuperscript{3}, Birendra Chhotaray\textsuperscript{2,4},  Arghajit Jana\textsuperscript{2} and P. C. Agrawal\textsuperscript{5}}  
\affilOne{\textsuperscript{1}National Space Institute, Technical University of Denmark, Elektrovej 327-328, DK-2800 Lyngby, Denmark\\}
\affilTwo{\textsuperscript{2}Astronomy and Astrophysics Division, Physical Research Laboratory, Navrangapura, Ahmedabad - 380009, Gujarat, India\\}
\affilThree{\textsuperscript{3}School of Physics and Technology, Wuhan University, Wuhan 430072, China\\}
\affilFour{\textsuperscript{4}Indian Institute of Technology, Gandhinagar - 382355, Gujarat, India\\}
\affilFive{\textsuperscript{5}Department of Astronomy and Astrophysics (Retired), Tata Institute of Fundamental Research, Colaba, Mumbai - 400005, India\\}

\twocolumn[{

\maketitle

\corres{gaurava@space.dtu.dk}
\msinfo{*}{*}

\begin{abstract}
We present the results obtained from analysis of two {\em AstroSat} observations of the high mass X-ray binary pulsar OAO~1657-415. The observations covered 0.681--0.818 and 0.808--0.968 phases of the $\sim$10.4 day orbital period of the system, in March and July 2019, respectively. Despite being outside the eclipsing regime, the power density spectrum from the first observation lacks any signature of pulsation or quasi-periodic oscillations. However, during July observation, X-ray pulsations at a period of 37.0375~s were clearly detected in the light curves. The pulse profiles from the second observation consist of a broad single peak with a dip-like structure in the middle across the observed energy range. We explored evolution of the pulse profile in narrow time and energy segments. We detected pulsations in the light curves obtained from 0.808--0.92 orbital phase range, which is absent in the remaining part of the observation. The spectrum of OAO~1657-415 can be described by an absorbed power-law model along with an iron fluorescent emission line and a blackbody component for out-of-eclipse phase of the observation. Our findings are discussed in the frame of stellar wind accretion and accretion wake at late orbital phases of the binary. 

\end{abstract}

\keywords{stars: neutron --- pulsars: individual: OAO~1657--415 --- X-rays: stars.}
}]

\doinum{12.3456/s78910-011-012-3}
\artcitid{\#\#\#\#}
\volnum{000}
\year{0000}
\pgrange{1--}
\setcounter{page}{1}
\lp{1}

\section{Introduction}

\source is an accreting high mass X-ray binary pulsar, discovered with Copernicus satellite in 1978 (Polidan et al. 1978). The X-ray pulsations from the neutron star were detected at 38.2~s using HEAO~A-2 observations (White \& Pravdo 1979; Parmar et al. 1980). Later, the system was identified as an eclipsing binary by using 1991 and 1992 observations with the Burst and Transient Source Experiment (BATSE) onboard Compton Gamma Ray Observatory (CGRO; Chakrabarty et al. 1993). This study also revealed the orbital period of the system to be P$_{orb}$ = 10.4 days along with the eclipse duration of about 1.7 days. Despite dedicated efforts in the optical band, no counter part was detected up to  a limitings magnitude of V$<$23. A highly reddened B-type supergiant star was subsequently  discovered as the optical companion of the pulsar within {\em Chandra} X-ray error box (Chakrabarty et al. 2002). The spectral class of the donor star was refined to be a Ofpe/WNL type star which is thought as a transitional object between the main sequence and Wolf-Rayet stars (Mason et al. 2009; Mason et al. 2012). Observed X-ray eclipses and massive nature of the companion established the binary system as an eclipsing high mass X-ray binary. The source distance is measured to be 4.4--12 kpc (Chakrabarty et al. 2002; Mason et al. 2009), consistent with the measurement of 7.1$\pm$1.3 kpc provided by Audley et al. (2006). The recent $Gaia$ data suggests a relatively lower  distance of 2.2$^{+0.5}_{-0.7}$~kpc based on the parallax angle of the optical companion (Malacaria et al. 2020).

In most of the high mass X-ray binary (HMXB) systems, a neutron star resides as the compact object. The mass accretion onto the neutron star in these systems takes place either via stellar wind accretion, Be-disk accretion, or Roche-Lobe overflow from the optical companion (see e.g. Reig
2011; Walter et al. 2015). Based on the mass accretion mechanisms and the nature of donor star, the HMXBs can be classified into Be-X-ray binaries (BeXBs) and supergiant X-ray binaries (SGXBs). The compact object in BeXBs co-rotates around a non-supergiant companion of class III-V, and accretes directly from a circumstellar disk of the Be star. On the other hand, the SGXBs consist of a super-giant optical companion of luminosity class I-II. The mass accretion in SGXB systems occurs through the stellar wind accretion (wind-fed accretion) or via an accretion disk formed after the capture of stellar wind or Roche-Lobe overflow (disk-fed accretion). Some of the SGXBs are also known to be eclipsing systems due to edge-on periodic obscuration of the compact object by the super-giant companion. Typical X-ray luminosity of the neutron star in SGXBs is in the range of 10$^{34}$--10$^{37}$ \fluxcgs (Mart{\'\i}nez-N{\'u}{\~n}ez et al. 2017). The source luminosity can also vary by a factor of 5-100 within a ks time-scale due to flaring activities (see, e.g. {F{\"u}rst} et al. 2010, Naik, Paul \& Ali 2011, Jaisawal et al. 2020). 

Various sub-classes of HMXBs represent a unique position in spin period vs. orbital period diagram (Corbet diagram; Corbet 1986). The BeXB systems show a strong correlation between the orbital and spin periods, whereas the wind-fed SGXBs are distributed in a horizontal line. There are, however, three disk-fed SGXBs viz. Cen~X-3, SMC~X-1 and LMC~X-4, that exhibit an anti-correlation between the orbital and spin periods in the Corbet diagram. In exception to the known pattern, OAO~1657-415 occupies an intermediate position in the Corbet diagram, between the wind-fed and disk-fed SGXBs (Chakrabarty et al. 1993; Jenke et al. 2012). 

Since discovery, the pulse period of \source changes stochastically at a rate of $\dot{\nu}$ $\sim$ 8.5$\times$10$^{-13}$~Hz~s$^{-1}$ (Baykal 1997; Bildsten et al. 1997; Baykal 2000). Steady spin-up and spin-down patterns, as seen in Cen~X-3, are also observed in the system (Bildsten et al. 1997). The changes in the spin episodes, however, can not be explained by the theory of torque reversal without considering the formation of a transient accretion disk (Baykal 1997). Based on a positive correlation between X-ray luminosity and the spin frequency during a spin-down phase in 1997, the presence of a prograde disk was suggested (Baykal 2000). On investigation of almost two decades of long-term spin evolution with  BASTE and  Gamma-Ray Burst Monitor (\fermi/GBM) observations, two accretion modes are inferred in this system (Jenke et al. 2012). The first mode is due to the disk-wind accretion (formation of an accretion disk after stellar wind accretion) where a stable accretion disk produces a correlation between flux and spin-up of the neutron star. In the latter case, a direct stellar wind accretion produces almost no correlation between the flux and spin-down parameter of \source at a lesser accretion rate (Jenke et al. 2012). Recently, Kim \& Ikhsanov (2017) proposed  a magnetic levitating disk hypothesis to explain the spin evolution in \source.

Being located at a low galactic latitude, \source is highly absorbed with a column density of 10$^{23}$~cm$^{-2}$ (Polidan et al. 1978, Kamata et al. 1990). The energy spectrum of the pulsar can be described by a power law continuum with an exponential high energy cutoff along with a soft excess and prominent emission lines at 6.4, 6.7, and 7.1~keV (Kamata et al. 1990; Audley et al. 2006; Barnstedt et al. 2008; Pradhan et al. 2014; Jaisawal \& Naik 2014; Pradhan, Raman \& Paul 2019).  {\em ASCA} observations were performed on 1994 March 22 and 1997 September 17 between orbital phase ranges of -0.001--0.074 and -0.21--0.11 (mid-eclipse time as phase zero), respectively (Audley et al. 2006). Presence of a dust scattered X-ray halo in \source was  suggested from these data. As expected, the observed source flux was very low near the mid-eclipse region with a dominant 6.4 keV line compared to the 6.7 keV line in the data from the first {\em ASCA} observation. The second {\em ASCA} observation covered entire eclipse along with out-of-eclipse phases before and after eclipse. The energy continuum in this phase was found to be absorbed along with detection of 6.4 and 7.1 keV iron emission lines (Audley et al. 2006). 

\suzaku observation of the source in September 2011, covering 0.12--0.34 orbital phase range, revealed flaring activities in the soft and hard X-ray light curves (Jaisawal \& Naik 2014; Pradhan et al. 2014). Detailed time-resolved spectroscopy of the \suzaku data suggested the accretion of clumpy material as the cause of flare-like episodes during the observation. Strong 6.4 and 7.1 keV lines were also detected. These lines  mostly originated from the neutral and ionized iron atoms within the accretion radius of 19 lt-sec (Jaisawal \& Naik 2014). Using {\em BeppoSAX} observation, a presence of a cyclotron absorption line was suggested at 36 keV (Orlandini et al. 1999). However, later studies with {\em INTEGRAL} and {\em Suzaku} did not confirm the feature (Barnstedt et al. 2008; Pradhan et al. 2014; Jaisawal \& Naik 2014).

 In the present paper, we study the properties of the source by using two \astrosat observations, carried out at mid and late orbital phases of the orbit (with mid-eclipse time as phase zero) in 2019 . Data analysis is presented in Section~2, followed by timing and spectral results in Section~3 and 4. The discussion and conclusion are summarized in Section~5.

\section{Observations and Data Analysis}

\astrosat is the first Indian multi-wavelength astronomical satellite launched by Indian Space Research Organization on 28 September 2015 (Agrawal 2006, Singh et al. 2014). It provides a broad-band coverage from optical to X-ray bands for exploring the nature of the cosmic sources. There are five sets of instruments such as Soft X-ray Telescope (SXT; Singh et al. 2017), Large Area X-ray Proportional Counters (LAXPCs; Agrawal et al. 2017, Antia et al.  2017), Cadmium Zinc Telluride Imager (CZTI; Rao et al.  2017), a Scanning Sky Monitor (SSM; Ramadevi et al.  2018), and Ultraviolet Imaging Telescope (UVIT; Tandon et al.  2017), onboard the satellite. In this paper, we study two observations of \source with SXT and LAXPC instruments. These observations were performed on  31 March and 4 July 2019, covering orbital phase ranges of 0.681--0.818 and 0.808--0.968, respectively. The orbital phase range is calculated based on the ephemeris of the binary system provided by Jenke et al. (2012). In our study, the CZTI data from both the epoch of observations are not used as the source was faint for the detector. The UVIT was not operational during these epochs. The log of the observations are given in Table~1.

\begin{table}
\caption{Log of observations of \source with {\it AstroSat}.}
\begin{tabular}{ccccc}
\hline
\hline
ObsID       &Start Date     &Expo.  &$\phi_{orb}$ \\
	        &(MJD)		     &(ks)  &\\
\hline

{\tt XX}\_9000002824   &58573.50     &59.8  &0.681--0.818 \\     
{\tt XX}\_9000003012   &58668.85     &62.2  &0.808--0.968 \\
 \hline
\hline
\end{tabular}
\label{log}
\tablenotes{Here {\tt XX} stands for A05\_205T01. $\phi_{orb}$ represents the orbital phase of the binary system. }
\end{table}

\begin{figure}[!bt]
\centering
\includegraphics[height=3.8in, width=3.3in, angle=0]{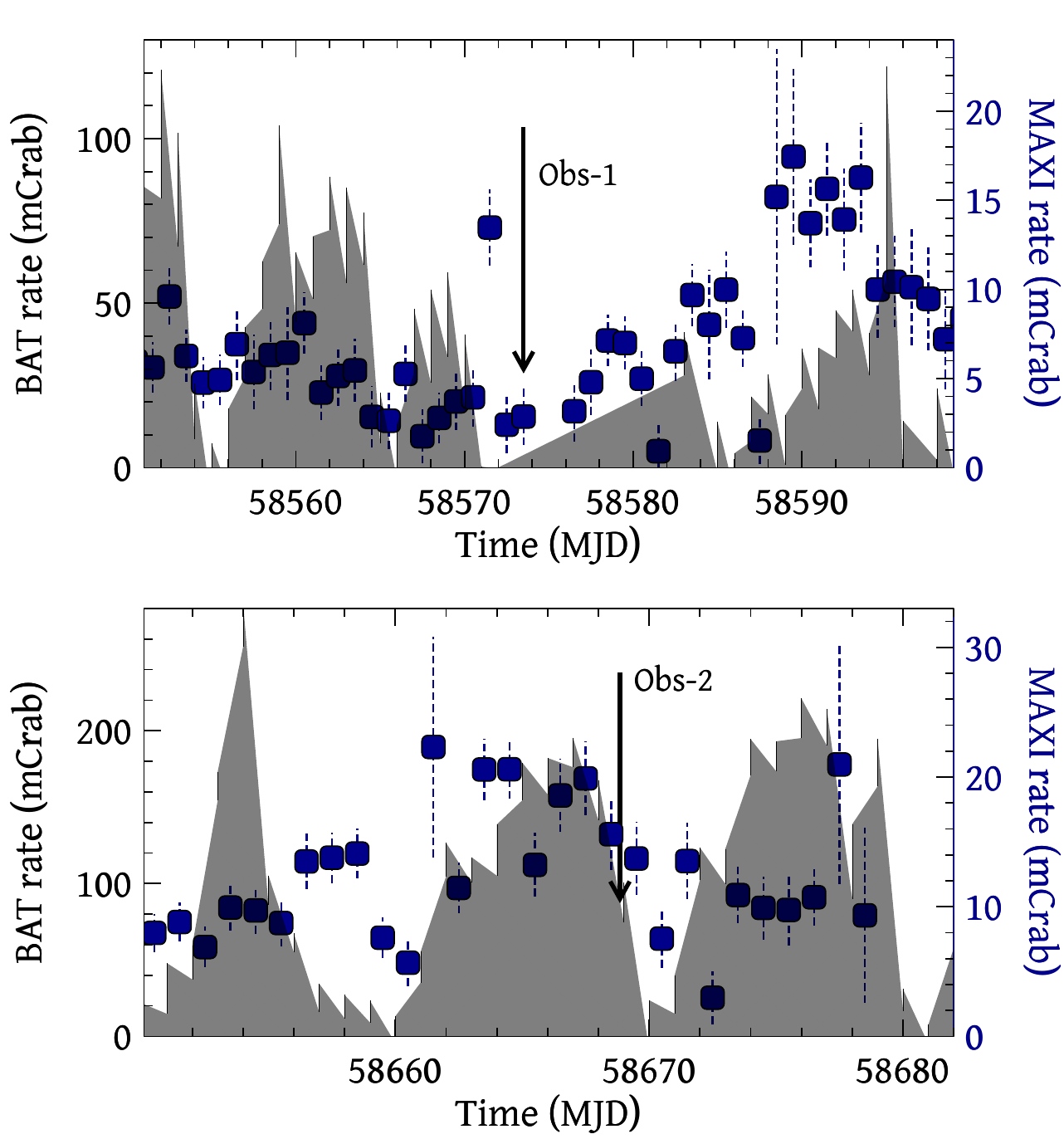}
\caption{Long term monitoring light curves of \source with MAXI (blue) and BAT (shaded) in 2-20 keV and 15-50 keV ranges, respectively. Arrows in both the panels represent the dates of the \astrosat observations. The pulsar appears to be weak during the first \astrosat observation (top panel). However, the second observation caught the source while changing from a relatively bright to faint phase of the binary orbit (bottom panel).}
\label{maxi-swift}
\end{figure}


\begin{figure*}[bt!]
\begin{center}$ 
\begin{array}{cc}
 \includegraphics[height=3.55in, width=3.35in, angle=0]{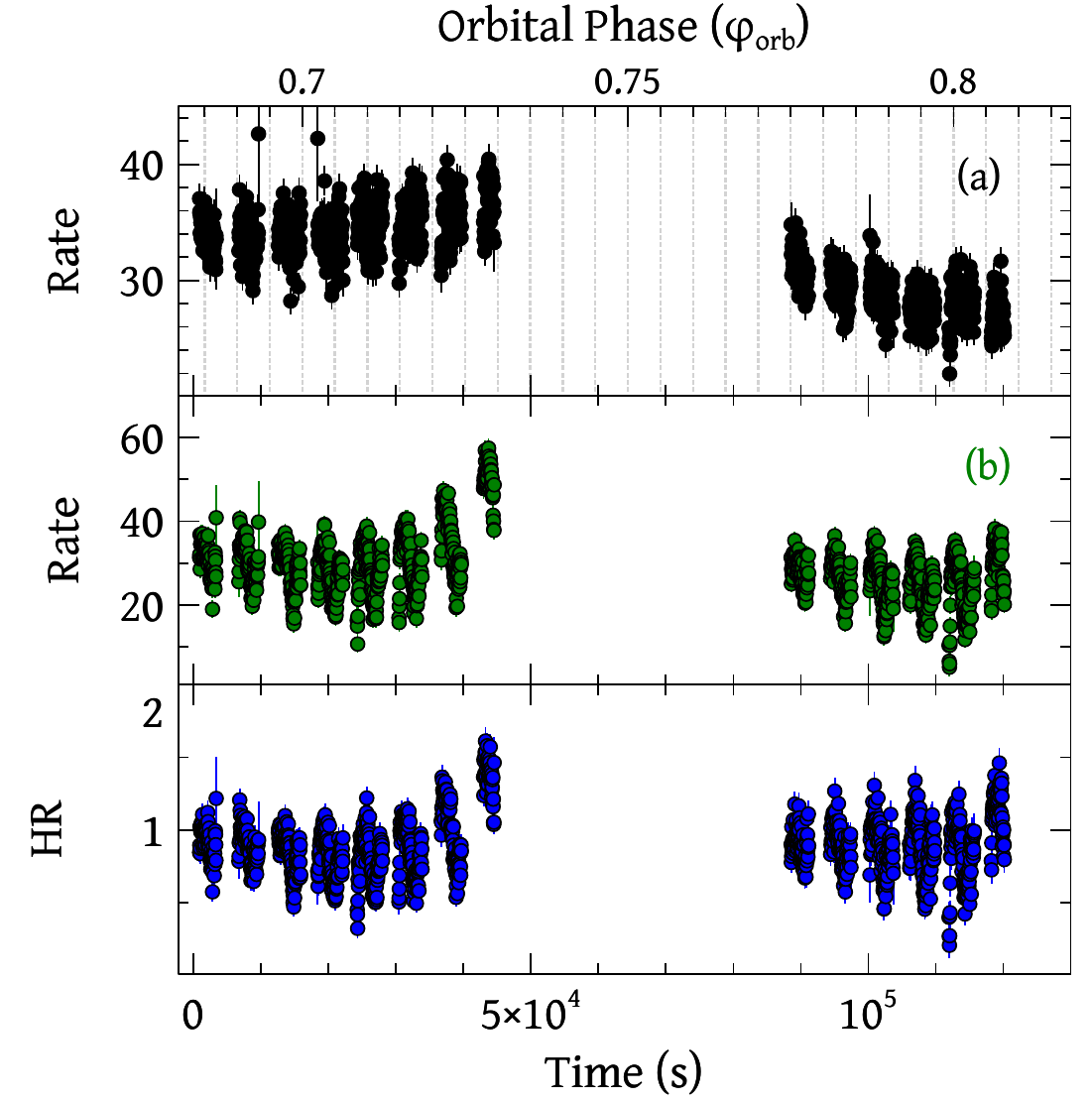} 
 \includegraphics[height=3.55in, width=3.35in, angle=0]{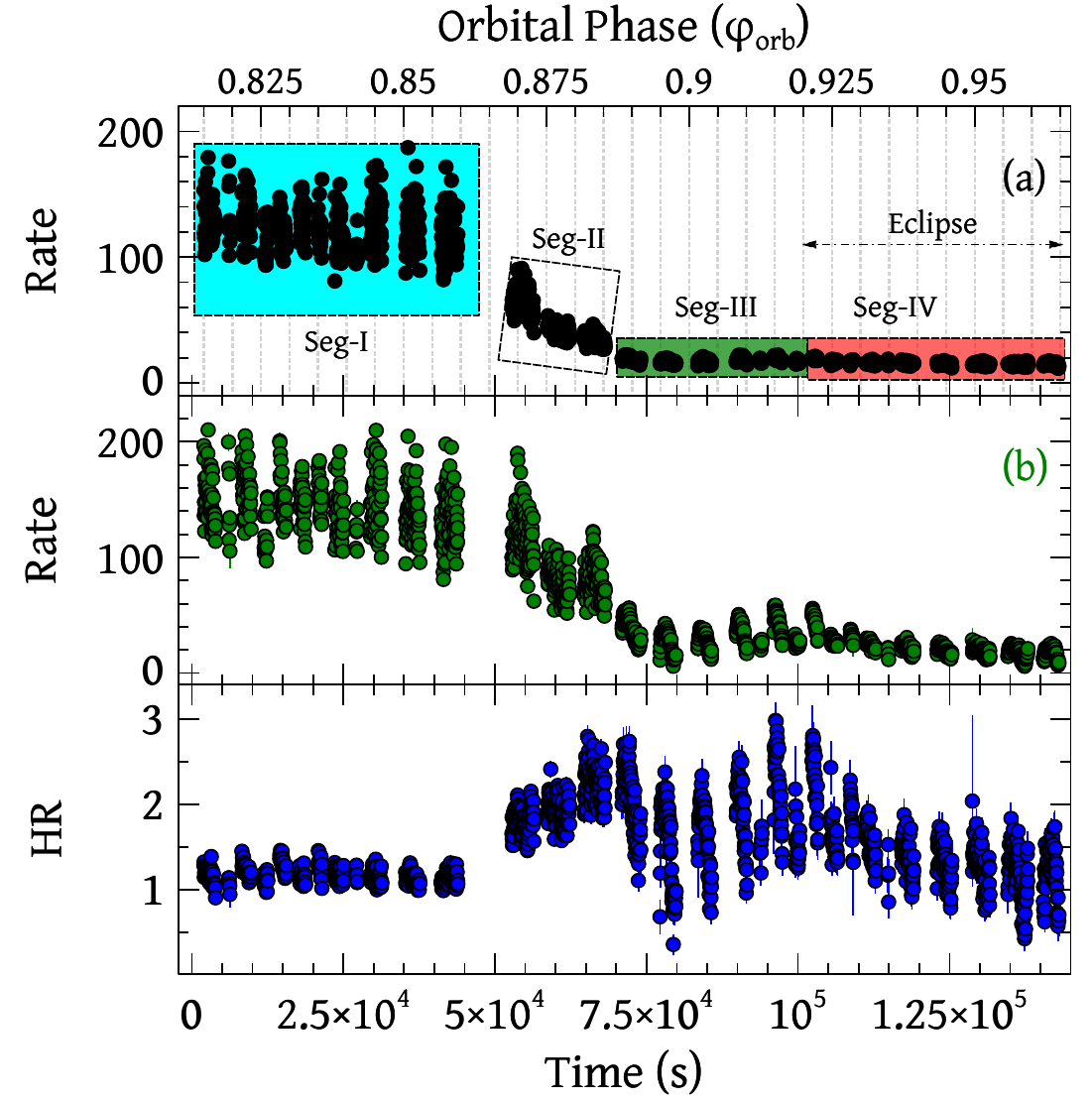} 
 \end{array}$
 \end{center}
\caption{Light curves from first and second {\em AstroSat}/LAXPC20 observations of \source on 31 March 2019 and 4 July 2019 are shown in left and right panels, respectively. The top and middle panels on both sides represent light curves in 3--10 keV and 10--80 keV ranges, respectively. The hardness-ratio, the ratio between light curves in 10--80 keV (middle panel) and 3--10 keV (top panel) energy bands, are shown in the bottom panels. The data from second observation are divided into four segments for further analysis and represented with different colors (top panel of the right side of the figure).  The y-axis in top and middle panels represents the source rate in counts per second unit.}
\label{lc-obs}
\end{figure*}


The SXT is a soft X-ray focusing telescope consisting of a CCD detector and sensitive in 0.3--8 keV energy range. The effective area and energy resolution of SXT is 128 cm$^2$ and 5--6\% at 1.5 keV and 22 cm$^2$ and 2.5\% at 6 keV, respectively. The observations of \source were carried out with SXT operating in photon counting mode, yielding a time resolution of 2.4~s. We used standard pipeline for SXT data reduction and merging tool {\tt sxtevtmergertool} provided by the \astrosat Science Support Cell (ASSC\footnote{\url{http://astrosat-ssc.iucaa.in/}}). As the pulsar was faint for SXT during the first observation, the data were used for spectral analysis only. The source spectrum was extracted from a 5~arcmin circular region centered at the source coordinate on the SXT chip using {\tt XSELECT} package. On the other hand, the light curves and spectra were extracted from the second observation by considering a source circular region of 5~arcmin. The background spectrum was obtained from a source free region on the SXT chip. 

The three LAXPCs are sensitive to X-ray photons in the 3--80 keV range and provide a total effective area of 8000 cm$^2$ at 15 keV. The time and energy resolution of the LAXPC units are 10~$\mu$s and 12\% at 22~keV, respectively. During both the observations, data from LAXPC20 were considered in our analysis. The data from LAXPC10 and LAXPC30 units were not used due to the presence of high background and gain issues with the instrument during observations (Antia et al. 2017). Using the standard data analysis routines ({\tt LAXPCsoftware}), the event mode data are analyzed to obtain the source light curves and spectral products. 
The LAXPC background products are obtained from the observation  using standard routines, recommended by the team. A systematic uncertainty of 2\% is also added in the LAXPC spectrum.


\begin{figure}[!bt]
 \begin{center}$
 \begin{array}{c}
 \includegraphics[height=3.2in, width=2.4in, angle=-90]{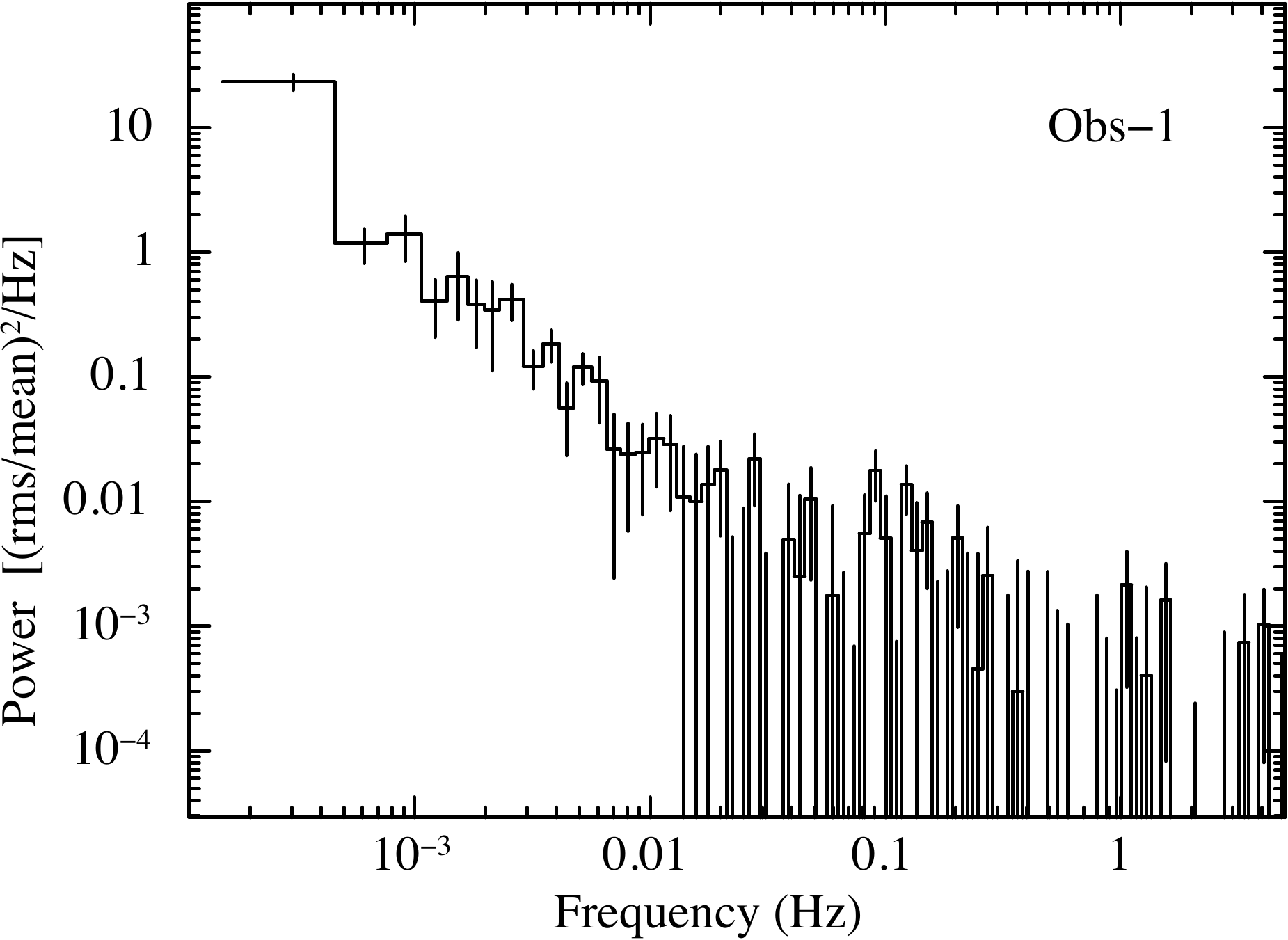} \\ 
 \end{array}$
 \end{center}
\caption{Power density spectrum of \source obtained from the light curves in 3-80 keV range from the LAXPC20 data of first \astrosat observation. Absence of peaks corresponding to the spin period of the pulsar can be seen. }
\label{pds1}
\end{figure}


\begin{figure*}[bt!]
 \begin{center}$
 \begin{array}{ccccc}
 \includegraphics[width=0.35\textwidth, angle=-90]{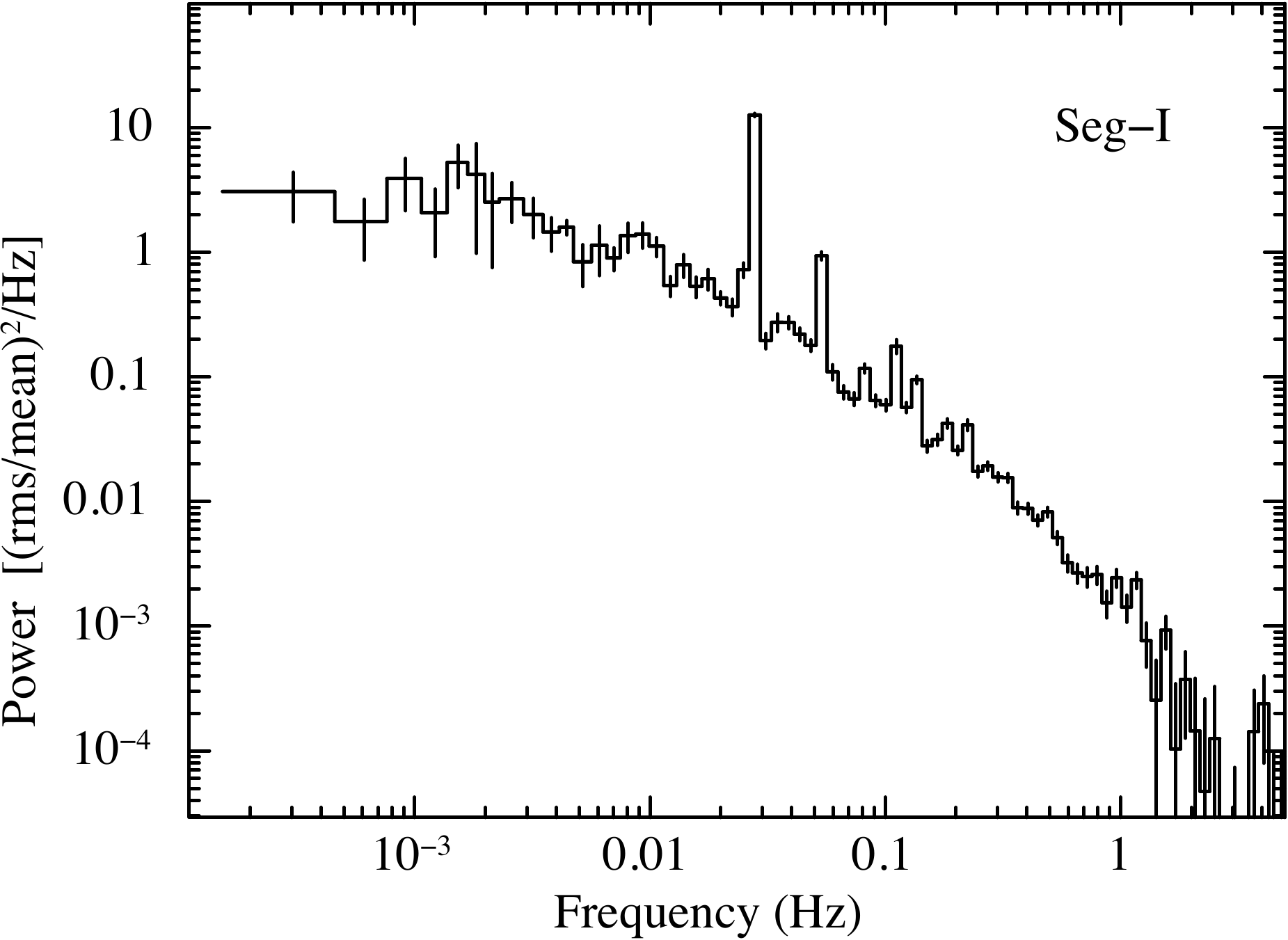} &
  \includegraphics[width=0.35\textwidth, angle=-90]{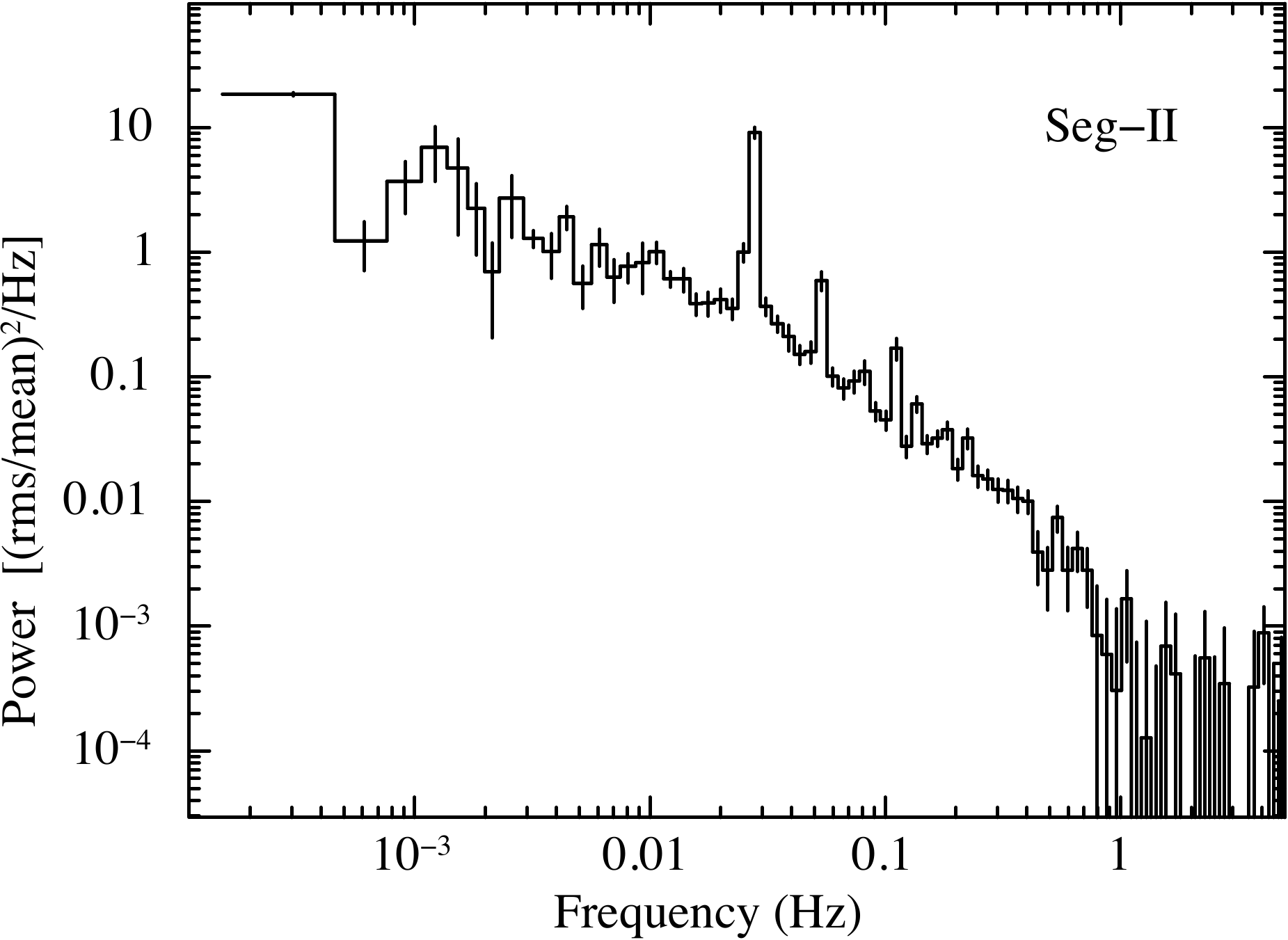} \\
  \includegraphics[width=0.35\textwidth, angle=-90]{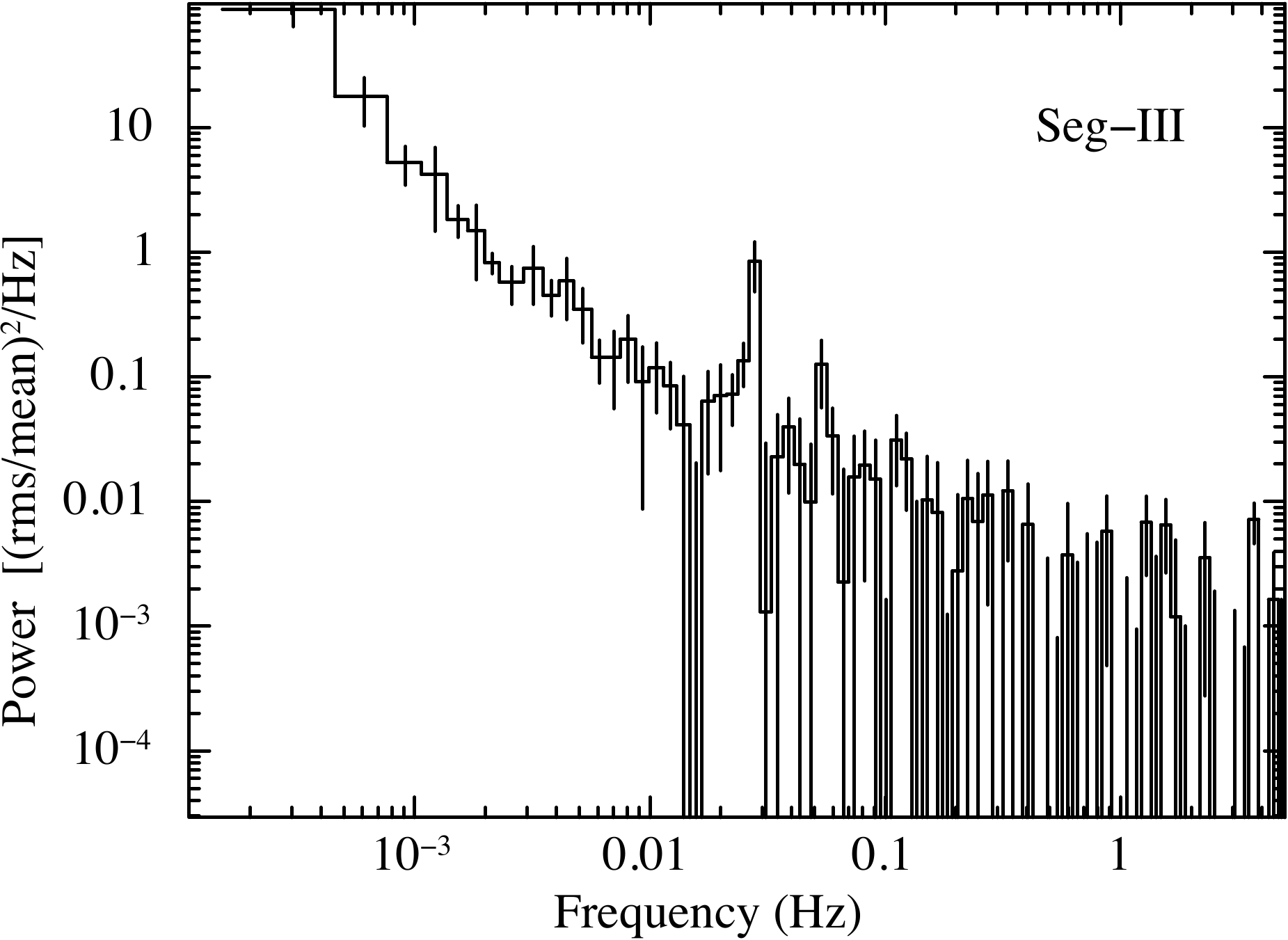} &
    \includegraphics[width=0.35\textwidth, angle=-90]{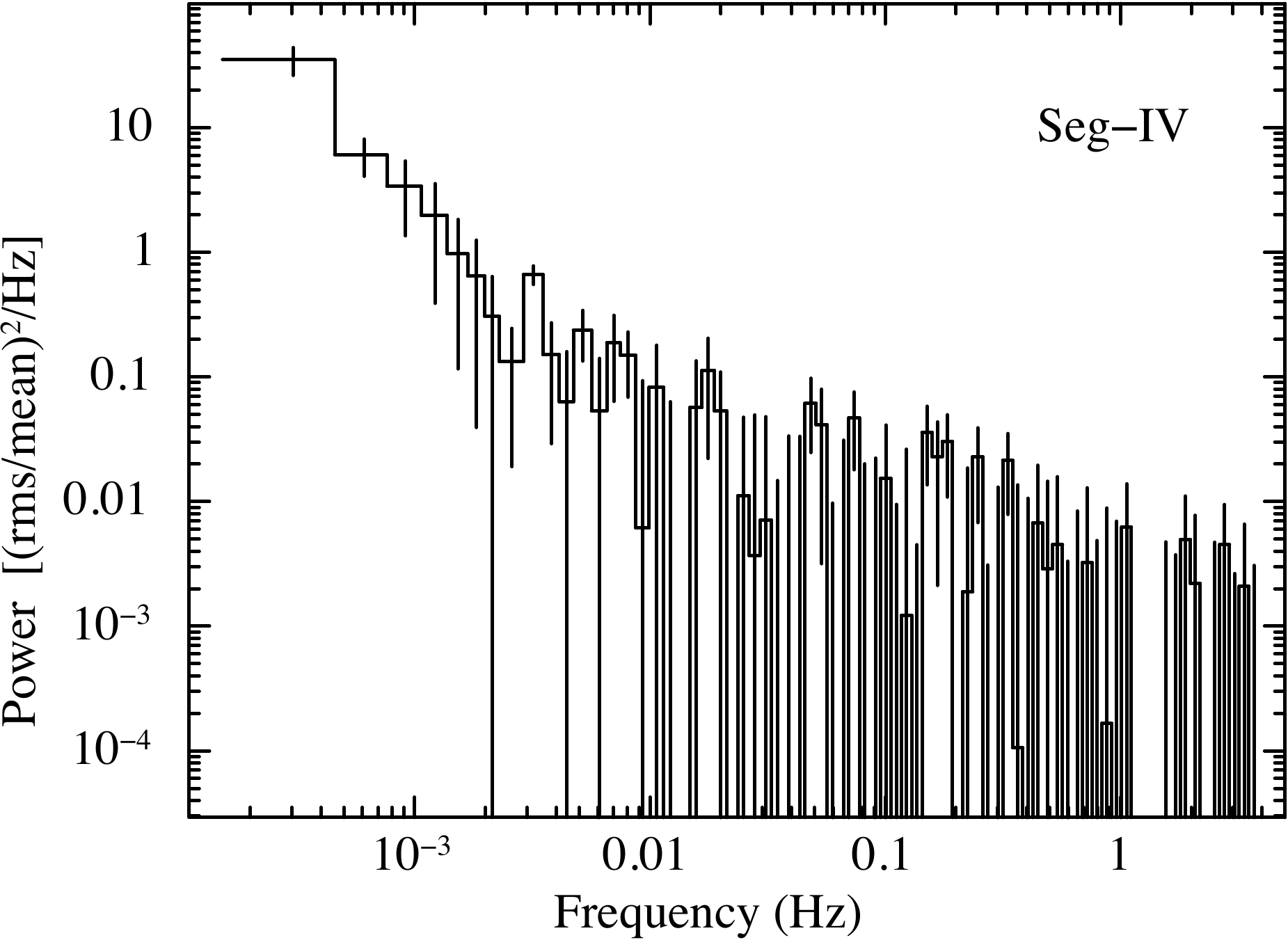} \\
 \end{array}$
 \end{center}
\caption{Power density spectra (PDS) of \source obtained from the segmented light curves of second \astrosat observation in 3-80 keV range. Presence of peaks at frequency corresponding to the spin period of the pulsar and its harmonics can be seen in the first, second and third segments of the observation. These peaks are, however, absent in the PDS corresponding to the fourth segment (right bottom panel).}
\label{pds2}
\end{figure*}


\begin{figure*}[h!]
 \begin{center}$
 \begin{array}{ccccc}
 \includegraphics[height=1.95in, width=3.3in, angle=-90]{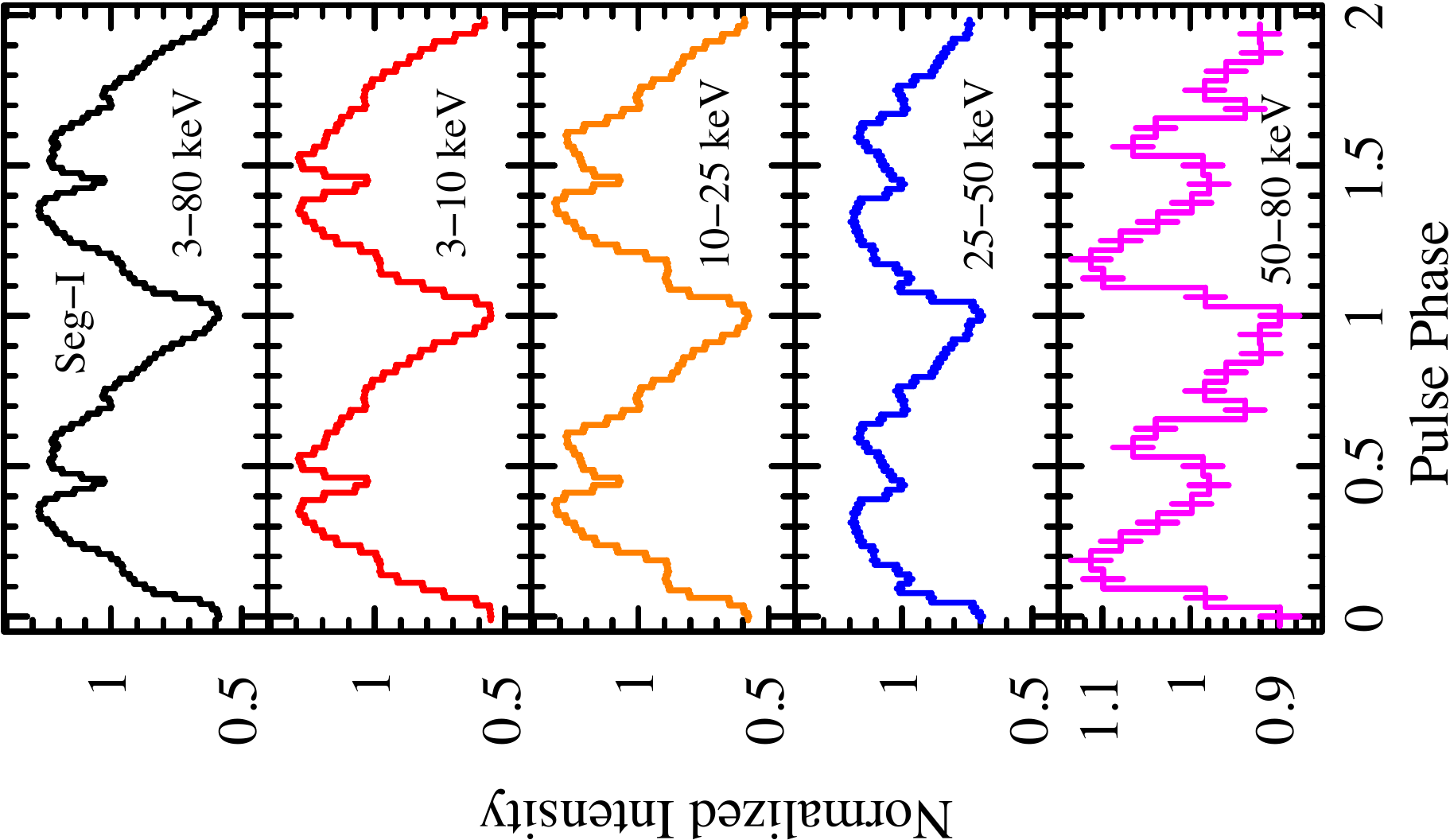} &
 \includegraphics[height=1.8in, width=3.3in, angle=-90]{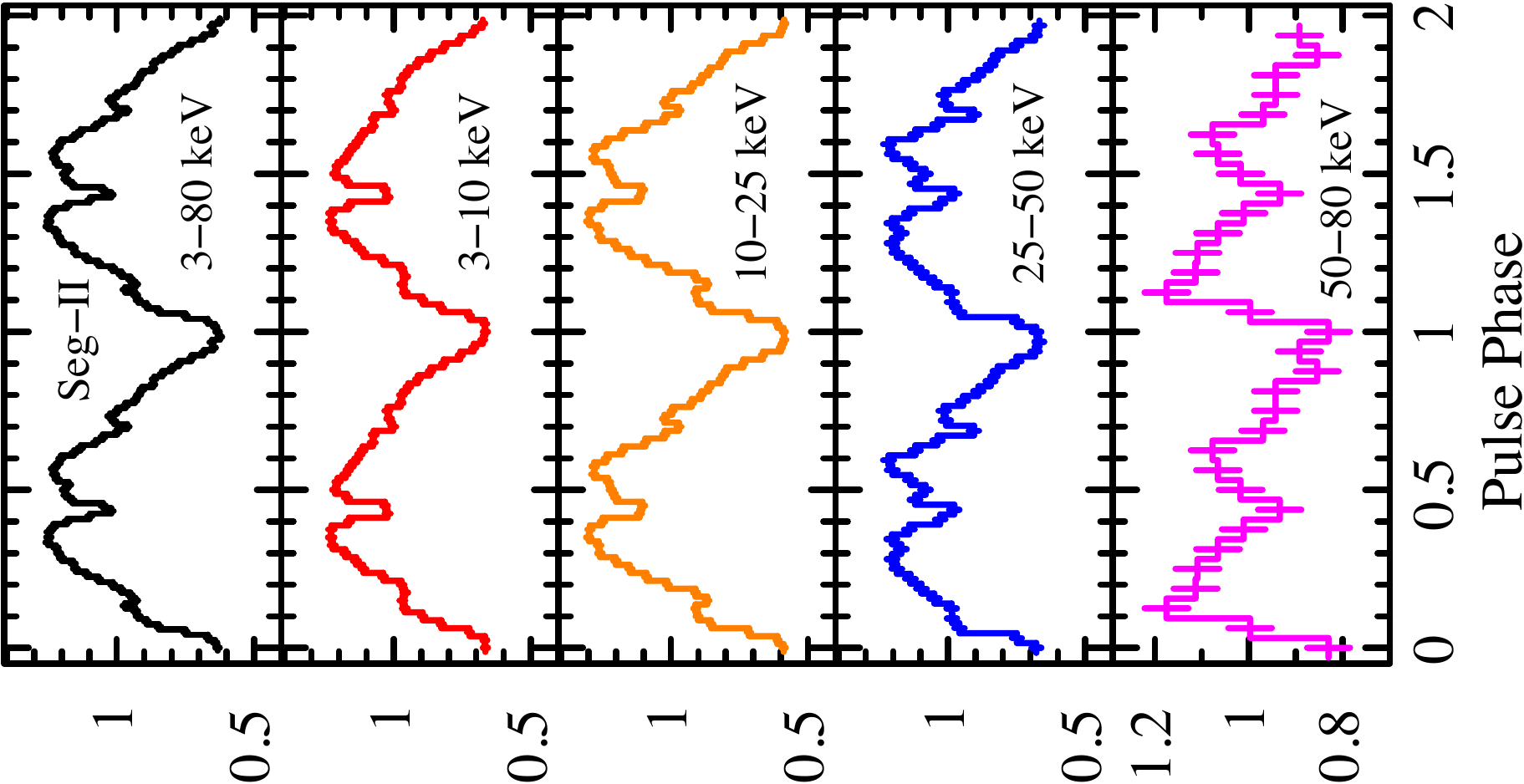} &
  \includegraphics[height=1.8in, width=3.3in, angle=-90]{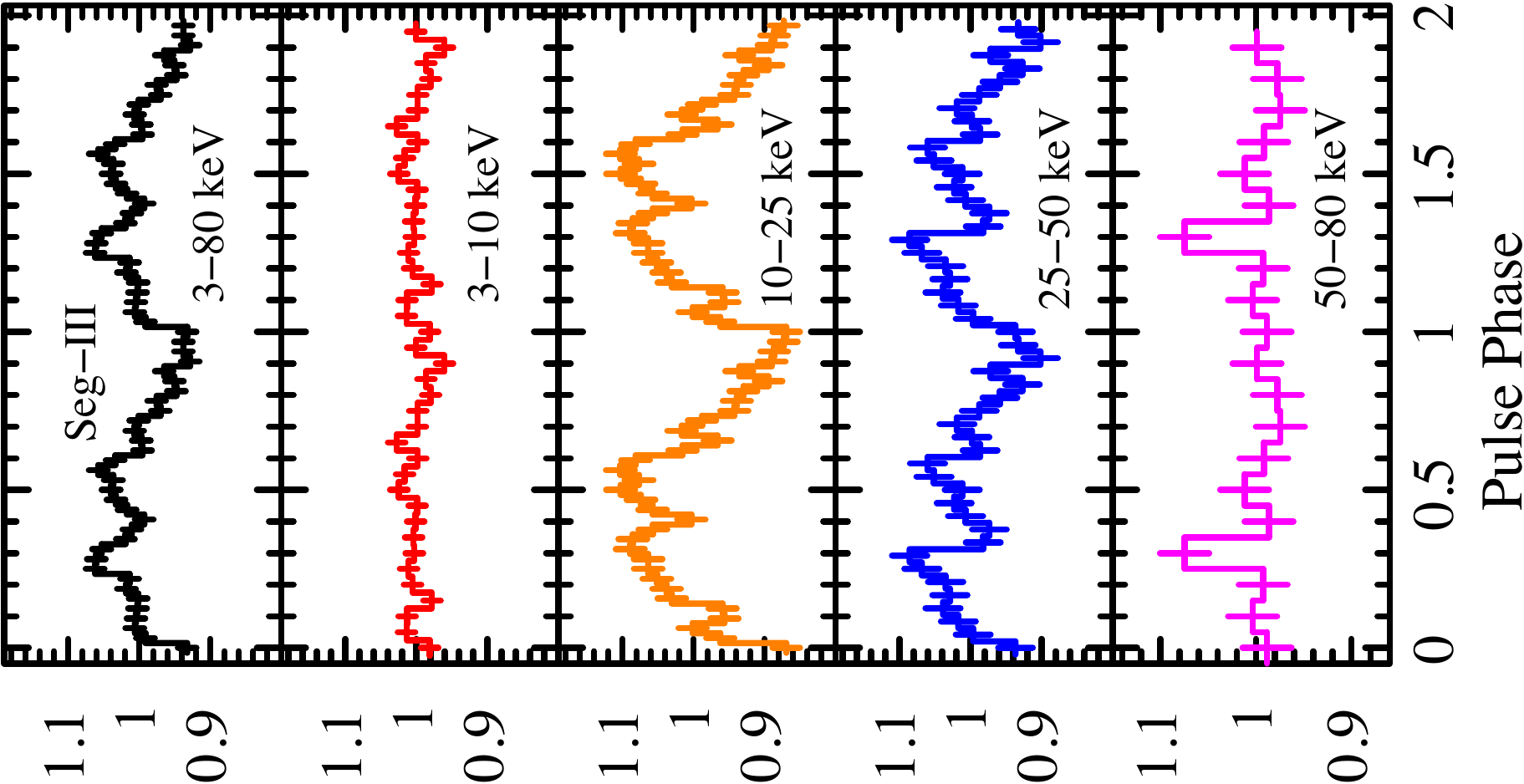} &
 \end{array}$
 \end{center}
\caption{Energy resolved pulse profiles of \source obtained by folding the light curves from LAXPC20 instrument during Seg-I, Seg-II and Seg-III of the second observation (as marked in top right panel of Figure~\ref{lc-obs}) at the estimated spin period. Top panels show the pulse profiles of the pulsar in entire LAXPC energy range for different segments, whereas the other panels show the pulse profiles in narrow energy ranges (quoted in each panel). Two pulses are shown in each panel for clarity. The error-bars represent 1$\sigma$  uncertainties.}
\label{profile-seg}
\end{figure*}


\begin{figure}[t!]
\centering
\includegraphics[height=3.3in, width=2.4in, angle=-90]{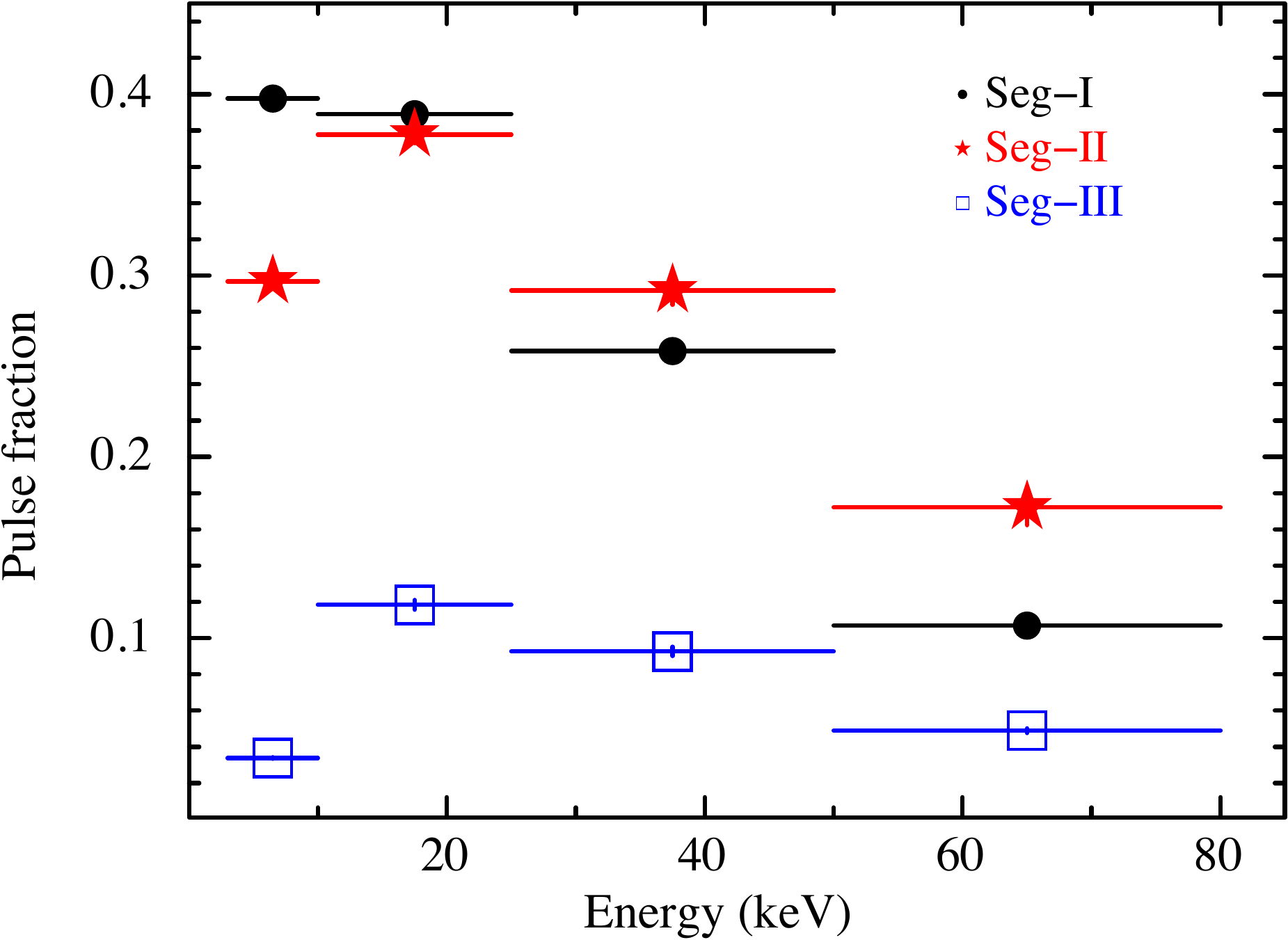}
\caption{Pulse fraction variation of the pulsar with energy obtained from pulse profiles in multiple energy bands.}
\label{pf}
\end{figure}

\section{Timing Studies}

\swift/BAT (Burst Alert Telescope, Krimm et al. 2013) and MAXI (Monitor of All-sky X-ray Image, Matsuoka et al. 2009) long term monitoring light curves of \source in 15--50 keV and 2--20 keV ranges, respectively, are shown in Figure~\ref{maxi-swift} to examine the overall activity of the pulsar during both the epochs of  \astrosat observations. During the first observation on 31 March 2019, \source was observed in a low intensity phase. However, during the second observation on 4 July 2019, the pulsar was observed to be brighter in the beginning and gradually entered the low flux level in the later part. The orbital phases covered during both the epochs of observations are 0.681--0.818 and 0.808--0.968 (Table~1). Background subtracted light curves obtained from LAXPC20 data of both the observations of the pulsar are shown in Fig~\ref{lc-obs}. The top panel (a) and middle panel (b) represent the source light curves in 3-10 keV and 10-80 keV energy ranges. The hardness ratio (HR), the ratio between the light curves in 10-80 keV and 3-10 keV ranges are also shown in the bottom panels of the figure. 

On comparison of light curves in 3--10 keV and 10--80 keV ranges (top two panels of both sides of Figure~\ref{lc-obs}), it can be noticed that the pulsar was in a low intensity phase during the first \astrosat observation, compared to that during the initial part of the second observation. During the first epoch of observation, the pulsar did not show any significant time variability in soft and hard X-ray light curves, despite being observed away from the eclipse regime. The hardness ratio was also found to be constant ($\sim$1). However, the second data set (right side of Figure~\ref{lc-obs}) shows a gradual evolution in the 3-10 keV and 10-80 keV light curves observed between 0.808--0.968 orbital phase of the binary. In the early part of this observation (out-of-eclipse phase), the pulsar was relatively brighter than that during the first \astrosat observation. The hardness ratio also found to change during the second observation. 

A search for X-ray pulsations was performed in the 3-80 keV barycentric corrected light curves, binned at 0.1~s, from both the observations. For this, the power-density spectra (PDS) were generated from the light curves using the Fast Fourier Transformation technique with the {\tt powspec} task of {\tt FTOOLS}. Absence of sharp narrow peaks in the PDS from the first observation (Figure~\ref{pds1}) suggests the non-detection of X-ray pulsations in the light curve of the pulsar. Pulsations were again searched in light curves in different energy bands such as 3-10 keV, 3-25 keV, and 10-80 keV ranges. However, we failed to detect any pulsating signal from the neutron star during the first observation.  The second observation of \source was carried out by covering out-of-eclipse and eclipse phases of the binary orbit. For the pulsation search, the entire observation was divided into four different segments (Seg-I, II, III, \& IV) on the basis of source intensity. These segments are marked in different colors in the top right panel of Figure~2. It should be noted that the fourth segment (Seg-IV) represents the duration of the eclipse of the neutron star during the second \astrosat observation. The PDS obtained from the 3-80 keV segmented light curves show signatures of strong pulsations along with its harmonics for Seg-I, II, \& III (Fig.~\ref{pds2}). The PDS obtained from Seg-IV, however, did not show any such signature at a frequency corresponding to the spin period of the pulsar (right bottom panel of Fig.~\ref{pds2}). This suggests that pulsations are present in the light curves during the out-of-the eclipse phase of the pulsar and absent during the eclipsing phase.

We, then, applied the chi-square maximization technique (Leahy 1987) using {\tt efsearch} task to determine the pulsation period. From the PDS analysis and the chi-square maximization technique, the spin period of the pulsar is estimated to be 37.0375(8)~s from the out-of-eclipse phases (Seg-I, II, \& III) of the second \astrosat observation. We checked the long term spin frequency history of \source using \fermi/GBM\footnote{\url{https://gammaray.nsstc.nasa.gov/gbm/science/pulsars.html}} data. The GBM instrument did not detect any pulsation in the source from MJD 58571.37 to MJD 58574.35 during which the first \astrosat observation was carried out. However, pulsations were detected during the second \astrosat observation as well as with \fermi/GBM. It is also worth to note that the count rate observed in the first observation is relatively higher than the same in third segment of second observation. The presence of pulsations at lower source count rate in the third segment confirms that the source was not into the propeller regime during the first observation.

 The background subtracted light curves in 3--80 keV energy range from the LAXPC20 data for Seg-I, II, \& III were folded at the estimated spin period of the pulsar. The pulse profiles from these segments are shown in the top panels of Figure~\ref{profile-seg} (left to right). A single peaked profile with a dip at the peak is observed in first three segments. Non-detection of pulsations in the light curve from Seg-IV, as the neutron star entered into the eclipsing phase of the binary (Figure~\ref{lc-obs}), pulse profiles for this segment were not generated. 

Evolution of the pulse profile with energy was investigated by generating profiles in different energy ranges. The light curves were extracted in several energy ranges such as 3-10 keV, 10-25 keV, 25-50 keV, and 50-80 keV, for Seg-I, II, \& III. These light curves were folded with the estimated spin period and presented in Figure~\ref{profile-seg} in second, third, fourth, and fifth panels (from top to bottom) for respective segments. In our study, pulsations are effectively detected up to 80 keV in first and second segments. The dip in the peak of the profile was also found to present up to higher energies.  The third segment which is in 0.89-0.92 orbital phase range (out-of-eclipse phase), appears like the quiescent phase of the pulsar (Figure~2, top right panel). Significant decrease in pulsar intensity during this segment, affected the soft X-ray pulse profile more. A single peaked structure with a dip appears only in the hard X-rays above 10 keV. The pulsation is  detected up to $\sim$50 keV in this case. 

To quantify the nature of these pulsating components, pulse fraction from the pulse profiles of \source is calculated and shown in Figure~\ref{pf}. In our study, the pulse fraction is defined as the ratio between the difference and sum of maximum and minimum intensities observed in the pulse profile. The pulse fraction from Seg-IV is not estimated as the pulsation was not detected in this segment. From the figure, a decreasing trend in the value of pulse fraction with energy, from a maximum of $\sim$40\%,  was seen for Seg-I and II. It suggests that the fraction of lower energy X-ray photons contributing towards pulsation is higher compared to the hard X-ray photons. In contrast, a marginally increasing trend between the pulse fraction and energy was seen for the Seg-III. Though a decreasing trend in the pulse fraction with energy can be clearly seen for all segments, the pulse fraction values in 3-10 keV range for Seg-II and Seg-III are found to be less than corresponding values in 10-25 keV ranges, whereas for Seg-I, it is comparable. This can be due to the fact that the soft X-ray photons are subjected to absorption by the particles in the interstellar medium. The gradual decrease in the values of pulse fraction with energy is anticipated when the source gets obscured by dense matter in the form of clumps of stellar wind, accretion wake, or the binary companion (e.g. in Seg-III). Pulse profiles of the pulsar were also generated from the SXT data. However, as the source is heavily absorbed in soft X-ray ranges, the profiles were not suitable to draw any meaningful information and not shown here.

\begin{figure}[t!]
\centering
\includegraphics[height=2.7in, width=3.2in, angle=0]{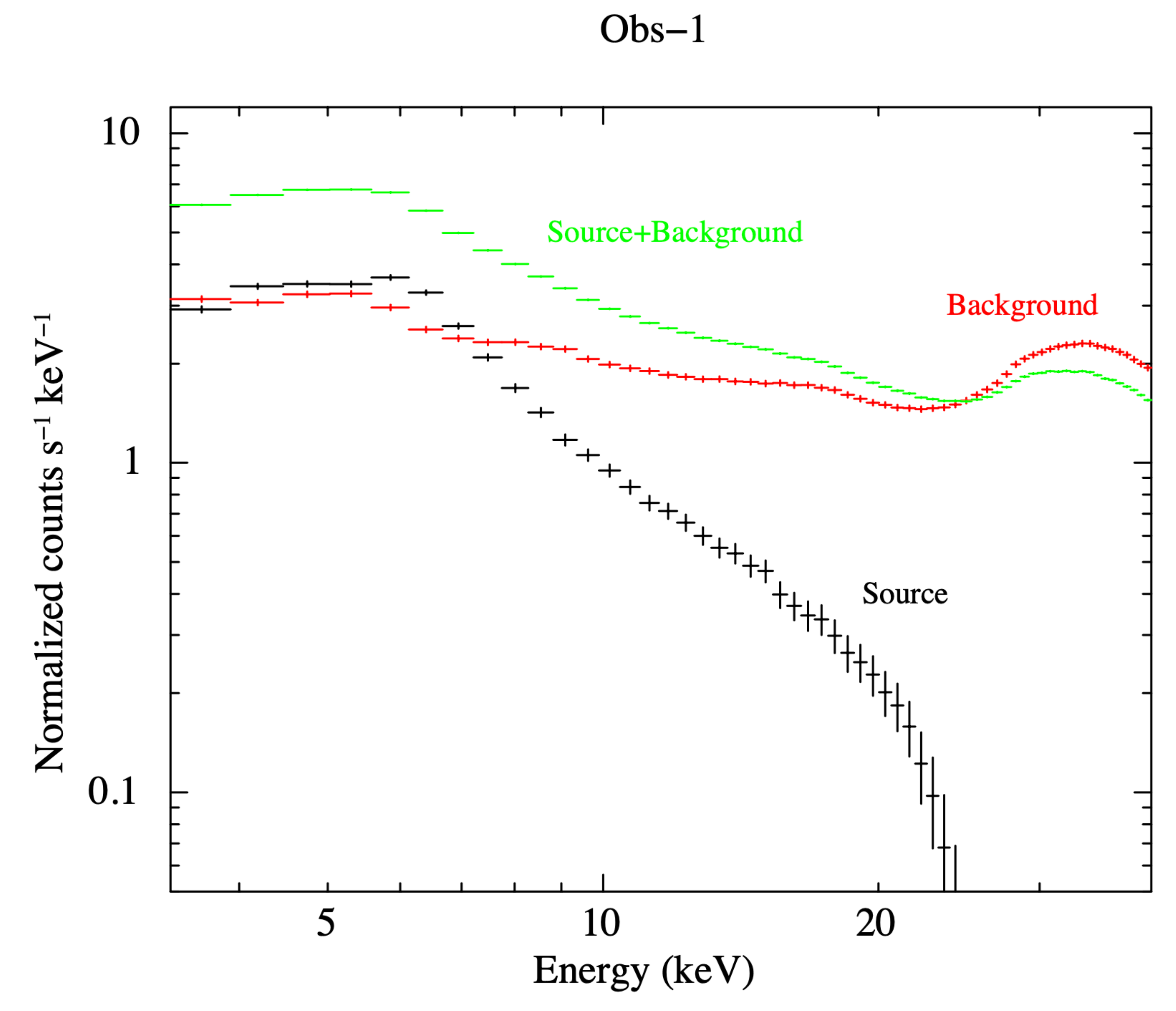}
\caption{Comparison between source and background energy spectra obtained from the first  \astrosat observation of \source.}
\label{spectrum_obs1}
\end{figure}

\begin{figure}[bt!]
 \begin{center}$
 \begin{array}{cccc}
 \includegraphics[height=3.3in, width=2.5in, angle=-90]{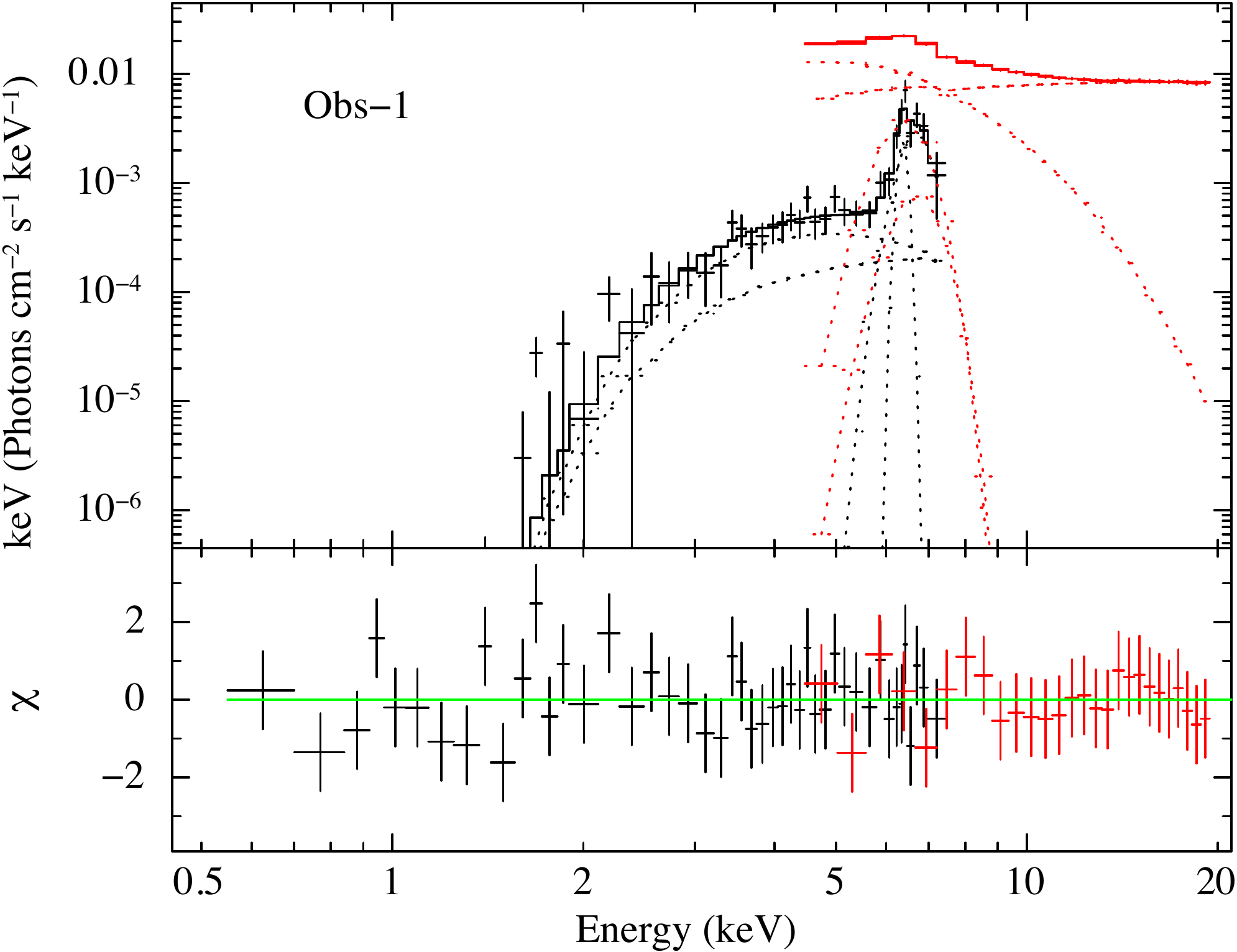}  \\
 \includegraphics[height=3.3in, width=2.5in, angle=-90]{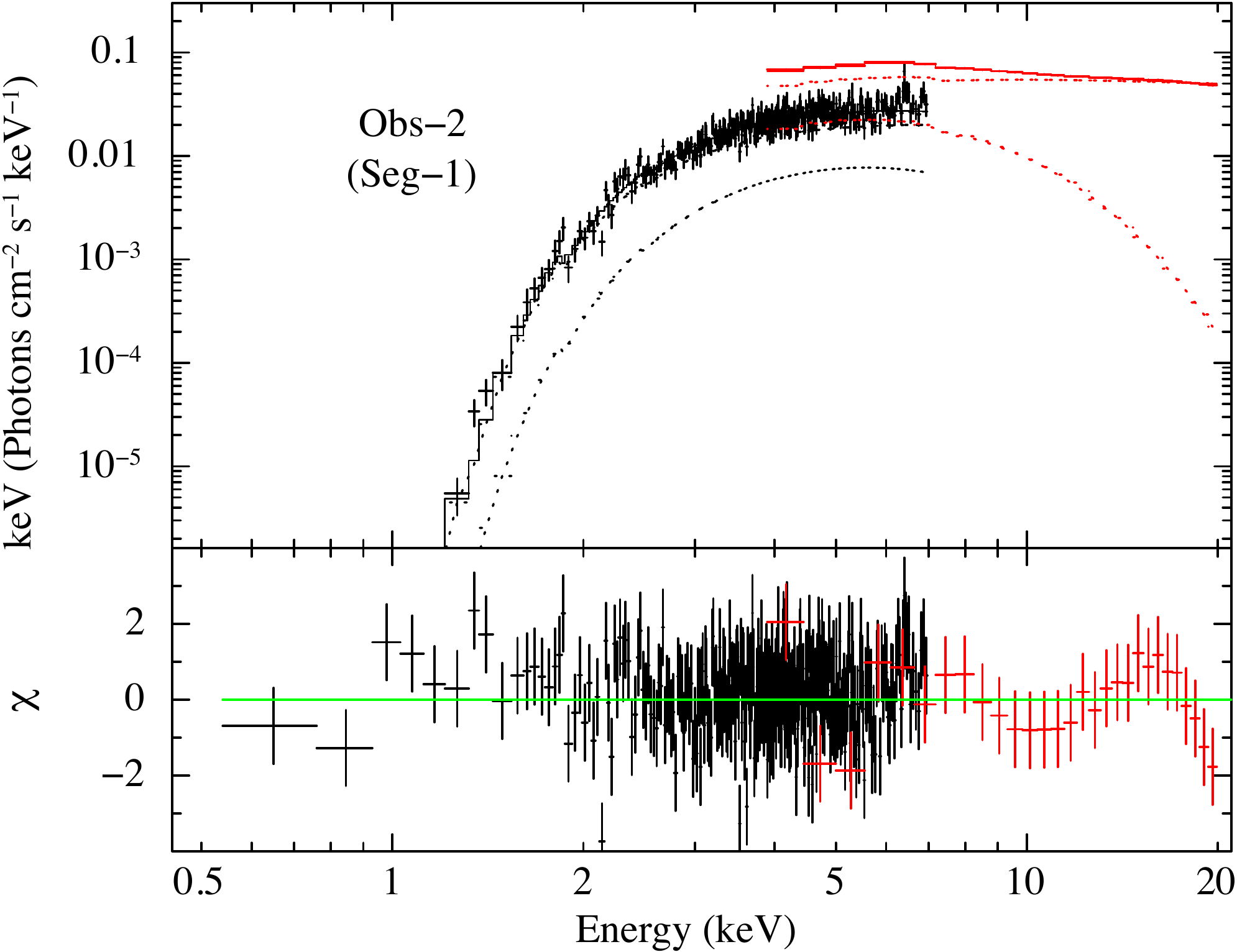} \\
  \includegraphics[height=3.3in, width=2.5in, angle=-90]{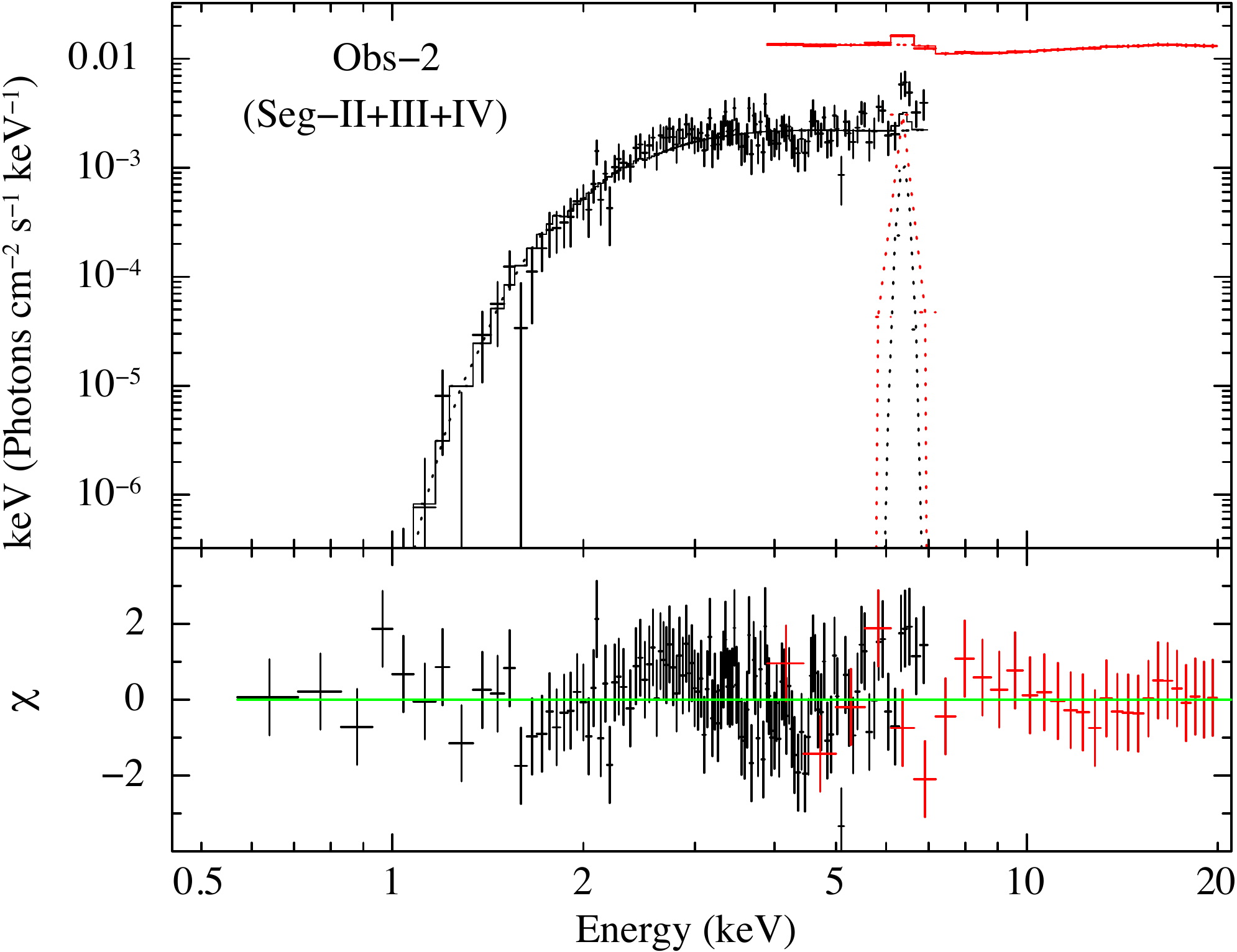} \\
 \end{array}$
 \end{center}
\caption{Best-fitting energy spectra of \source during the first (top panel) and second (middle and bottom panels) \astrosat observations of the source. The data from SXT and LAXPC instruments in 0.5-7 and 3.5-20 keV are used in the spectral fitting, respectively. }
\label{spec}
\end{figure}  

\section{Spectral Studies}

To understand the cause of non-detection of pulsations in the first \astrosat observation and the intensity variation during the second observation, we carried out spectral analysis by using SXT and LAXPC20 data from both the observations. The spectral fitting package {\tt XSPEC}  of version 12.10.0 (Arnaud 1996) was used.  In the beginning, we extracted observed source+background and background spectra using {\tt laxpc\_make\_spectra} and {\tt laxpc\_find\_back} tasks of {\tt LAXPCsoftware} package, respectively, from LAXPC20 data of Obs-1. The observed, background, and source spectra of the source from LAXPC20 of Obs-1 are shown in Figure~\ref{spectrum_obs1}. It can be seen that the source spectrum is limited up to $\sim$22 keV. As we aim to carry out our spectral analysis using data from both the observations, we restricted ourselves to fit the data up to 20 keV in our fitting.

In our analysis, we considered 0.5-7 keV spectrum from SXT and 3.5-20 keV spectrum from LAXPC20. The 0.5-20 keV energy spectrum obtained from the first observation was fitted with several standard models such as power law, high energy cutoff power law, power law with blackbody component etc., along with the Galactic absorption column density {\tt TBabs} (Wilms, Allen \& McCray 2000). Only an absorbed power law with a blackbody component ({\tt bbodyrad} in {\tt XSPEC}) was able to fit the spectrum better, though excess residuals were observed in 6-7 keV range. This excess was resolved into two iron emission lines at 6.4 keV and 6.7 keV while fitting SXT data alone. Similar multiple iron emission lines have also been seen in other accretion powered X-ray pulsars such as GX~1+4 (Naik, Paul \& Callanan 2005; Yoshida et al. 2017), Cen~X-3 (Naik, Paul \& Ali, 2011), Swift~J0243.6+6124 (Jaisawal et al. 2019). In the present study, these lines could not be distinguished in the LAXPC data due to relatively poor spectral resolution of the instrument. Therefore, the parameters of the Gaussian components for emission lines were fixed at values obtained from fitting SXT data alone, in the joint fitting of SXT and LAXPC data.  Because of poor energy resolution of LAXPC,  these parameters were not allowed to vary. This resulted in improving the goodness of fit per degree of freedom to $\chi^2_\nu$=$\chi^2/\nu$ $\approx$1. The best-fit spectral parameters obtained from our fitting are quoted in Table~2.  The iron line parameters quoted in Table~2, such as line energy, width, equivalent width and line flux are obtained by fitting the SXT data alone.

The LAXPC20 light curve of the pulsar during the second \astrosat observation shows different intensity phases of the source (Figure~2). During the normal phase (Seg-I), the source intensity was maximum which gradually decreased (Seg-II) to low intensity phases (Seg-III \& IV). As the source flux was extremely low during Seg-III \& IV and the duration of Seg-II is very short, spectral analysis was carried out by considering Seg-I and added Seg-II, III \& IV, separately. While fitting the SXT and LAXPC spectra for Seg-I, the absorbed power-law model with a blackbody component provided acceptable fit. However, there was no clear signature of iron emission line in the residuals. While fitting the spectra corresponding to added Seg-II, III, and IV, a simple absorbed power-law model or an absorbed power-law with blackbody model did not fit the data well. Considering the low intensity nature of the neutron star at late orbital phases, a partial covering component was tried along with the absorbed power-law model. Addition of partial covering component to the absorbed power-law model improved the fitting. We also detected a 6.4 keV iron fluorescence line in the spectra for this added segment. The best-fitted spectral parameters obtained from the spectral fitting of data from both the \astrosat observations are given in Table~2, whereas the energy spectra along with corresponding residuals are shown in Figure~\ref{spec}. We used {\tt cflux} convolution model for flux estimation in our study.

\begin{table}[t!]
\tabularfont
\centering
\caption{Best-fitting spectral parameters (with 90\% errors) of \source.}
\begin{tabular}{ |l | c|cc|}
\hline

 Parameters                      &  Obs-1  &\multicolumn{2}{c}{Obs-2}\\
                              &  &Seg-I     &II+III+IV     	 \\	
\hline
N$_{\rm H1}$$^a$         &15$\pm$3   &6.6$\pm$0.5    &4.9$\pm$0.4  \\
N$_{\rm H2}$$^a$         &--     &--      &166$\pm$29\\
CF            &--     &--      &0.56$\pm$0.06\\
$\Gamma$            &1.1$\pm$0.2 &1.2$\pm$0.1  &1.6$\pm$0.1             \\
norm (10$^{-4}$)          &2.5$\pm$1  &341$\pm$1  &172$\pm$1\\
kT (keV)                &1.2$\pm$0.1  &1.7$\pm$0.1   &-- \\
BB$_{\rm norm}$    &0.2$\pm$0.1   &1.4$\pm$0.4    &-- \\
                   &              &               &\\
\underline{Iron Lines} &          & &\\
E1 (keV)      &6.37$\pm$0.1    &--   &6.4$\pm$0.1\\
W1 (keV)     &0.2$\pm$0.2   &--  &0.1\\
Flux1$^b$   &2.3$\pm$1  &--     &1$\pm$0.6 \\
Eqw1 (keV)  &1.1$\pm$0.7  &--  &0.12$\pm$0.07 \\
        &              &               &\\
E2 (keV)     &6.68$^{+0.3}_{-0.1}$    &--  &--\\
W2  (keV)    &0.4$^{+0.3}_{-0.4}$   &-- &-- \\
Flux2$^b$   &1.9$\pm$1  &--  &-- \\
Eqw2 (keV)  &0.8$\pm$0.4  &--  &-- \\
               &              &               &\\
Flux$^c$   	  &1.6$\pm$0.3        &55$\pm$2  &13$\pm$3 \\
$\chi^2_\nu$ ($\nu$)      &0.77 (61)   &1.03 (310)   &1.01 (142)    \\
\hline
\end{tabular}
\label{log}
\tablenotes{$^a$: in 10$^{22}$~cm$^{-2}$ unit; $^b$: unabsorbed line flux in 10$^{-12}$~\fluxcgs; \\
$^c$: 0.5-30 keV unabsorbed flux in 10$^{-11}$~\fluxcgs unit }
\end{table}

\section{Discussion and Conclusions}

We studied two \astrosat observations of \source carried out in March and July 2019 covering 0.681--0.818 and 0.808--0.968 phase ranges of the 10.4 day orbital period of the binary system, respectively.  During the first observation, no pulsation was detected in the SXT and LAXPC data. In this observation, despite being significantly away from the eclipse, the source was found weak in the soft and hard X-ray light curves. The 0.5-30 keV unabsorbed flux was estimated to be 1.6$\times$10$^{-11}$~\fluxcgs, corresponding to a source luminosity of $\approx$0.93$\times$10$^{34}$ and 9.4$\times$10$^{34}$ erg s$^{-1}$ at a distance of 2.2 and 7 kpc, respectively. The spectral analysis of the data from this observation revealed a high value of column density N$_{\rm H}$ of about 1.5$\times$10$^{23}$~cm$^{-2}$. Strong iron lines at 6.4 keV and 6.7 keV were also detected in the SXT spectrum obtained from the first observation. The equivalent width of the lines are estimated to be as high as 1 keV. Detection of such strong iron lines with high equivalent width suggests the presence of abundant material around the pulsar for reprocessing. A high value of equivalent width ($\sim$1 keV) is only possible when a dense absorbing medium faces directly the X-ray source (Inoue 1985). There are at least two possibilities of absorbing medium one can presume in front of the neutron star. First one, is an accretion wake. Using simulation study of accretion onto neutron stars in wind-fed sources, Blondin et al. (1990) suggested the formation of accretion wake at the late orbital phases. The density inside the non-steady accretion wake could be 100 times higher than the undisturbed stellar wind (Blondin et al. 1990). The dense wake can efficiently absorb the X-ray photons and hence reducing the source flux to lower values as observed in this case. Similar structure has been reported in case of 4U~1700--37 during the out-of-eclipse part of the binary between 0.63--0.73 phases (Boroson et al. 2003; Jaisawal \& Naik 2015). An increase in the column density is usually observed in these orbital phases. In contrast to the present study, Jaisawal \& Naik (2015) did not find any significant increase in the iron line parameters in case of 4U~1700--37. The study of \source in 0.12--0.34 orbital phase range with \suzaku revealed the presence of eclipse-like segment in the light curve. Time-resolved spectroscopy of the \suzaku data corresponding to 0.19--0.23 phase showed a high column density as well as strong 6.4 keV and 7.1 keV iron lines with equivalent width of about 1 keV and 0.3 keV, respectively (Jaisawal \& Naik 2014; Pradhan et al. 2014). The presence of a dense blob of material across the line of sight of neutron star was speculated within an accretion radius (Jaisawal \& Naik 2014). In the present study, the existence of a dense blob of material or the accretion wake can not be denied. If the material is dense, it can  absorb the X-ray photons, henceforth no pulsation is observed as seen during the first observation. Nonetheless, the detection of strong iron emission lines favors the hypothesis of absorption through a clumpy stellar wind in our study. 

Alternatively, the cessation of the pulsations can be possible in X-ray pulsars when the source enters into propeller regime (Illarionov \& Sunyaev 1975) at extremely low mass accretion rate  (see, e.g., in case of GX~1+4 and GRO J1744-28 reported by Cui 1997). If this is the case, the observed X-ray luminosity can be assumed as a limiting luminosity of the pulsar.  However, it is unlikely that the pulsar was in the propeller regime during first {\it AstroSat} observation. This is due to the detection of iron emission lines during this observation. 
Considering the presence of high column density and iron emission lines at late orbital phases of the binary orbit (first {\it AstroSat} observation), the disappearance of pulsations can be interpreted as due to the presence of accretion wake. 

From the second \astrosat data, pulsations were clearly detected in the out-of-eclipse data. The pulse profiles of \source were found to be singly peaked with a dip like structure at the peak of the profile. Usually this source is subjected to strong and dense stellar wind of the supergiant companion as it is in a close binary system. Inhomogenous distribution of stellar wind as well as clumpy wind accretion in SGXBs can affect the pulsed emission. The profile can be affected mostly in the soft X-rays due to  absorption. In the present and previous studies of OAO~1657--415, the presence of a dip in the profile is seen up to higher energies (Pradhan et al. 2014). Such type of profiles with absorption-like feature(s) at certain pulse phases are common in transient BeXRB pulsars (Maitra, Paul \& Naik 2012; Naik et al. 2013; Jaisawal, Naik \& Epili 2016). The complex shape of the pulse profiles in these Be/X-ray pulsars is interpreted as an effect of coupling between accreted material and magnetic field lines. 

We also carried out spectral investigation of the source during the second \astrosat observation by dividing the data into two parts. The first part of the data (Seg-I, Figure~2) was relatively brighter by a factor of 34 than the first \astrosat observation. We did not detect any iron emission line in the data from this segment. The SXT and LAXPC spectra from the second part of the second observation (added Seg-II, III \& IV, Figure~2), covering 0.87--0.968 orbital phase range, were  highly absorbed. Only a weak iron emission line at 6.4 keV was detected in this segment. As the spectral fitting is limited to 0.5-20 keV range, detection of the cyclotron absorption features (Jaisawal \& Naik 2017; Staubert et al. 2019) is not expected in the data.

In summary, we have studied two \astrosat observations in March and July 2019 at late orbital phases of the binary system. We did not detect any strong pulsation at the spin period of the pulsar in the first data set, observed in 0.681--0.818 phase range. The presence of relatively high column density and strong iron emission lines at 6.4 keV and 6.7 keV with an equivalent width of $\approx$1 keV, suggest the clumpy wind material close to the neutron star. A possibility of accretion wake can not be denied at a late orbital phases of wind-fed SGXBs. The pulsations are detected in the light curves obtained from the second observation. From the energy and time resolved pulse profiles, it is found that the pulsations are detected in the data before the source gets eclipsed by the optical companion. 

\vspace{-2em}

\appendix

\vspace{1.5em}

\section*{Acknowledgements}
We thank the anonymous referee for suggestions on our paper.
This publication uses the data from \astrosat
mission of the ISRO, archived at the Indian Space Science Data
Centre. We thank members of SXT, LAXPC, and CZTI instrument
teams for their contribution to the development of the instruments and analysis software. 
LAXPC data were processed by the Payload Operation Centre at TIFR, Mumbai. This work was performed utilizing the calibration data-bases and auxiliary analysis tools developed, maintained and distributed by the AstroSat-SXT team with members from various institutions in India and
abroad, and the SXT Payload Operation Center (POC) at the TIFR, Mumbai
(https://www.tifr.res.in/~astrosat\_sxt/index.html). SXT data were processed and verified by the SXT POC.
SN, BC and AG acknowledges
the support from Physical Research Laboratory which is funded
by the Department of Space, India.

\vspace{1em}




\begin{theunbibliography}{} 
\vspace{-1.5em}

\bibitem{latexcompanion} 
Agrawal P. C. 2006, Adv. Space Res., 38, 2989 
\bibitem{latexcompanion} 
Agrawal P. C., et al. 2017, J. Astrophys. Astron., 38, 30
\bibitem{latexcompanion} 
Antia H. M., et al. 2017, ApJS, 231, 10
\bibitem{latexcompanion} 
Arnaud K. A. 1996, in Jacoby G. H., Barnes J., eds, ASP Conf. Ser. Vol. 101,
Astronomical Data Analysis Software and Systems V. Astron. Soc. Pac., San Fransisco, p. 17
\bibitem{latexcompanion}
Audley M. D., Nagase F., Mitsuda K., Angelini L., Kelley R. L. 2006, MNRAS, 367, 1147
\bibitem{latexcompanion} 
Barnstedt J., Staubert R., Santangelo A., et al. 2008, A\&A, 486, 293
\bibitem{latexcompanion} 
Baykal A. 1997, A\&A, 319, 515
\bibitem{latexcompanion} 
Baykal A. 2000, MNRAS, 313, 637
\bibitem{latexcompanion} 
Bildsten L., Chakrabarty D., Chiu J., et al. 1997, ApJS, 113, 367
\bibitem{latexcompanion} 
Blondin J. M., Kallman T. R., Fryxell B. A., Taam R. E. 1990, ApJ, 356, 591
\bibitem{latexcompanion} 
Boroson B., Vrtilek S. D., Kallman T., Corcoran M. 2003, ApJ, 592, 516
\bibitem{latexcompanion} 
Chakrabarty D., Grunsfeld J. M., Prince T. A., et al. 1993, ApJ, 403, L33
\bibitem{latexcompanion} 
Chakrabarty D., Wang Z., Juett A. M., Lee J. C., Roche P. 2002, ApJ, 573, 789
\bibitem{latexcompanion} 
Corbet R. H. D. 1986, MNRAS, 220, 1047

\bibitem{latexcompanion} 
Cui W. 1997, ApJ, 482 163

\bibitem{latexcompanion} 
{F{\"u}rst} F., et al. 2010, A\&A, 519, 37

\bibitem{latexcompanion} 
Illarionov A. F., Sunyaev R. A. 1975, A\&A, 39, 185

\bibitem{latexcompanion} 
Inoue H. 1985, Space Sci. Rev., 40, 317
\bibitem{latexcompanion} 
Jaisawal G. K., Naik S. 2014, BASI, 42, 147
\bibitem{latexcompanion} 
Jaisawal G. K., Naik S. 2015, MNRAS, 448, 620
\bibitem{latexcompanion} 
Jaisawal G. K., Naik S., Epili P. 2016, MNRAS, 457, 2749
\bibitem{latexcompanion}
Jaisawal G. K., Naik S. 2017, in Serino M., Shidatsu M., Iwakiri W., Mihara T., eds, 7 years of MAXI: monitoring X-ray Transients, held 5-7 December 2016 at RIKEN, RIKEN, Saitama, Japan. p. 153
\bibitem{latexcompanion} 
Jaisawal G. K., et al. 2019, ApJ, 885, 18
\bibitem{latexcompanion} 
Jaisawal G. K., et al. 2020, MNRAS, 498, 4830
\bibitem{latexcompanion} 
Jenke P. A., Finger M. H., Wilson-Hodge C. A., Camero-Arranz A. 2012, ApJ, 759, 124
\bibitem{latexcompanion} 
Kamata Y., Koyama K., Tawara Y., et al. 1990, PASJ, 42, 785
\bibitem{latexcompanion} 
Kim V. Yu.,  Ikhsanov N. R. 2017, J. Phys.: Conf. Ser., 929 012005
\bibitem{latexcompanion} 
Krimm H. A., et al. 2013, ApJSS, 209, 14
\bibitem{latexcompanion} 
Leahy D. A. 1987, A\&A, 180, 275
\bibitem{latexcompanion} 
Maitra C., Paul B., Naik S. 2012, MNRAS, 420, 2307                         
\bibitem{latexcompanion} 
Malacaria C., Jenke, P., Roberts O. J., Wilson-Hodge C. A., Cleveland W. H., Mailyan B. 2020, ApJ, 896, 90
\bibitem{latexcompanion} 
Mason A. B., Clark J. S., Norton A. J., Negueruela I., Roche P. 2009, A\&A, 505, 281 
\bibitem{latexcompanion} 
Mason A. B., Clark J. S., Norton A. J., et al. 2012, MNRAS, 422, 199 
\bibitem{latexcompanion} 
Matsuoka, M., et al. 2009, PASJ, 61, 999
\bibitem{latexcompanion} 
Mart{\'\i}nez-N{\'u}{\~n}ez S. et al. 2017, Space Sci. Rev., 212, 59
\bibitem{latexcompanion} 
Naik S., Paul B., Callanan P.~J. 2005, ApJ, , 618, 866
\bibitem{latexcompanion} 
Naik S., Paul B., Ali Z. 2011, ApJ, 737, 79
\bibitem{latexcompanion} 
Naik S., Maitra C., Jaisawal G. K., Paul B. 2013, ApJ, 764, 158
\bibitem{latexcompanion} 
Orlandini M., dal Fiume D., del Sordo S., et al. 1999, A\&A, 349, L9 
\bibitem{latexcompanion} 
Parmar A. N., Branduardi-Raymont G., Pollard G. S. G., et al. 1980, MNRAS, 193, 49
\bibitem{latexcompanion} 
Pradhan P., Maitra C., Paul B., Islam N., Paul B. C. 2014, MNRAS, 442, 2691	
\bibitem{latexcompanion} 
Pradhan P., Raman G., Paul B. 2019, MNRAS, 483, 5687
\bibitem{latexcompanion} 
Polidan R. S., Pollard G. S. G., Sanford P. W., Locke M. C. 1978, Nature, 275, 296
\bibitem{latexcompanion} 
Ramadevi M. C. et al. 2018, J. Astrophys. Astron., 39, 11
\bibitem{latexcompanion} 
Rao A. R., Bhattacharya D., Bhalerao V. B., Vadawale S. V., Sreekumar S. 2017, Curr. Sci., 113, 595
\bibitem{latexcompanion} 
Reig P. 2011, Ap\&SS, 332, 1
\bibitem{latexcompanion} 
Singh K. P., et al. 2014, SPIE, 9144, 15
\bibitem{latexcompanion} 
Singh K. P. et al. 2017, J. Astrophys. Astron., 38, 29
\bibitem{latexcompanion}
Staubert R., et al. 2019, A\&A, 622, A61
\bibitem{latexcompanion}
Tandon S. N. et al. 2017, AJ, 154, 128
\bibitem{latexcompanion} 
Walter R., Lutovinov A. A., Bozzo E., Tsygankov S. S. 2015, A\&AR, 23, 2
\bibitem{latexcompanion} 
White N. E.,  Pravdo S. H. 1979, ApJ, 233, L121
\bibitem{latexcompanion} 
Wilms J., Allen A., McCray R. 2000, ApJ, 542, 914
\bibitem{latexcompanion} 
Yoshida Y., et al. 2017, ApJ, 838, 30 





\end{theunbibliography}
\end{document}